\documentclass{article}
\usepackage{amsmath, amsthm, amsfonts}
\usepackage[a4paper,margin=2.5cm]{geometry}
\newtheorem{thm}{Theorem}[section]
\newtheorem{cor}{Corollary}
\newtheorem{lem}{Lemma}
\newtheorem{pro}{Proof}
\newtheorem{propr}{Properties}
\newtheorem{prop}{Proposition}
\theoremstyle{definition}
\newtheorem{defn}[thm]{Definition}
\newtheorem{rem}{Remark}
\usepackage{graphicx}
 \usepackage{xcolor}
\usepackage{lineno}

\title{ The Alpha-Beta-Symmetric Divergence and their Positive Definite Kernels }
\author{Mactar Ndaw, Macoumba Ndour, and Papa Ngom\\
  \small LMA-Laboratoire de Math\'{e}matiques Appliqu\'{e}es\\
  \small Universit\'{e} Cheikh Anta Diop BP 5005 Dakar-Fann S\'{e}n\'{e}gal\\
  \small e-mail: mactarndaw1@gmail.com, macoumbandour@hotmail.fr, papa.ngom@ucad.edu.sn
}

\begin{document}
\maketitle

\begin{abstract}

 In the field of statistical modeling, the distance or divergence measure is a  criterion 
widely known and widely used tool for theoretical and applied statistical inference and data processing problems. 
In this paper, we deal with the  well-known Alpha-Beta-divergences (which we shall refer to as the AB-divergences) which are a family of cost functions parametrized by two hyperparameters and their tight connections with the notions of Hilbertian metrics and positive definite (pd) kernels on probability measures. An attempt is made to describe this dissimilarity measure, which can be symmetrized using its two tuning parameters, alpha and beta. We compute the degree of symmetry of the AB-divergence on the basis of Hilbertian metrics.
 We investegate the desirable properties that the proposed approach needs to build a positive definite kernel $\mathcal{K}(x,y)$ corresponding to this symmetric  AB-divergence. \\
 We establish the effectiveness of our approach with experiments conducted on Support Vector Machine (SVM) and the applicability of this method is described in an algorithm from this symmetric divergence in image classification.
 
 We perform experiments using the conditionally defined positive $\mathcal{K}$ and the kernel transformed $\mathcal{K}_t $  and show that these  kernels have the same proportion of errors for the Euclidean divergence and the Hellinger divergence. 
We also observe large r\'eductions in error for the Itakura-saito divergence with the K kernel  in classifications than classical Kernel methods.
\\

\end{abstract} 

\section{Introduction}

      Over the last few years, the need for specific design of kernels for a given data structure has been recognized by the kernel community. 
    Recently, a Hilbert space embedding for probability measures has been proposed, with applications including dimensionality reduction, independence testing and machine learning. 
      Therefore the use of specialized metrics and divergences measures in the successful design of dimensionality reduction techniques has been progressively acquiring much recognition. There are numerous real scenarios and applications for which the parameters of interest belong to non-flat manifolds, and where the Euclidean geometry results are unsuitable to evaluate the similarities. Indeed, this is usual case in the comparison of probability density functions. 
    So, for better results  the kernel should be adjusted as acurately as possible to the subjacent structure of the input space. 
  Kernel on probability measures are very handy for dealing with graph problems, trees, manifolds, acoustic and signal processing and they became very popular because of their many applications.
In probability theory, the distance between probability measures is used in studying pattern analysis see John Shawe Taylor and Mello Cristianini (2004).
Another application is in giving a bounded probability space $\mathcal{X}$ and  using the kernel to compare arbitrary sets in that space, by putting e.g the uniform measure on each set. This extremely useful to compare data of variable length,
 sequence data in bioinformatics for example, kernel methodes for predictting protein-protein in interactions (Asa Ben-Hur, Willian Stafford Noble 2005).\\

        The  20th century  witnessed  tremendous  efforts  to  exploit  new   distance/similarity   measures   for   a   variety   of applications.   There are a substantial number of distance/similarity measures encountered  in  many  different  fields  such  as  biology,  chemistry,  computer  science,  ecology,  information  theory,  geology,  mathematics,  physics,  statistics,  etc.  Distance or similarity measures are essential to solve many  pattern  recognition  problems  such
 as classification, clustering, and retrieval problems. Various distance/similarity measures that are applicable to compare two probability density functions.
 The advantages of discriminative learning algorithms and kernel machines are combined with generative modeling using
 a novel kernel between distributions. In the probability product kernel, data points in the input space are mapped to distributions over the sample space and a general inner product is then evaluated as the integral of the product of pairs of distributions. 
 Recently, developments in machine learning, including the emergence of support vector machines, have rekindled interest in kernel methods (Vapnik, 1998; Hastie et al., 2001) and take full advantage of well known probabilistic models. 
 These kernel methods have been widely employed to solve machine learning problems such as classification and clustering. Although there are many existing graph kernel methods for comparing patterns represented by undirected graphs, the corresponding methods for directed structures are less developed.
In particular for domains such as speech and images kernel functions have been suggested as good ways to combine an underlying generative model in the feature space and discriminant classifiers such as SVM. \\
Of  particular  concern  to  mathematicians  is  that  several divergence measures are  asymmetric.  \\
     However, in support vector machine classifier,  asymmetric kernel functions are not used so far, although they are frequently used in other kernel classifies. \\
   In this paper, we suggest an alternative procedure by exploiting  the symmetric AB-divergence measures, and present an information theoretic kernel method for assessing the similarity between a pair of directed graphs. \\ 
 In particular,  we show that our kernel method provides an efficient tool in  statistical  learning  theory,  and SVM  have  demonstrated  highly  competitive  performance  in  numerous real-world applications, such as medical diagnosis, bioinformatics, face recognition, image processing and text mining, which has established SVM as one of the most popular, state-of-the-art tools for knowledge discovery  and  data  mining.  Similar  to  artificial  neural  networks,  SVM possess the well-known ability of being universal approximators of any multivariate  function  to  any  desired  degree  of  accuracy.   \\

  \par

The remainder of this paper is organized as follows. 
In section 2, we set out the basic notations, the definitions and assumptions.
We will show the close relationship between Hilbert metrics and pd kernels so that in general, stratements for one category can be easily transferred to another. In section 3, we definit  a  Hilbertian 
metrics  the Alpha-Beta-Symmetric divergence (ABS-divergence) and the property of this divergence are studied and we are given a corresponding positive 
definite kernels. In section 4, the results of the simulations are presented. Therefore we evaluated the performance of the proposed metrics and kernels in tree classifications. And we proposed and apply a algorithm in experimental dataset to analysis the robustness of the divergence proposed. In the last section we presented the conclusion.

\section{Basic Notation and some results}

   For a class of Hilbertian metrics, that are metrics which can be isometrically embedded into a Hilbert space. We will also use the following function class to define this subclass of metrics.
\subsection{Hilbertian Metrics, Positive Definite Kernels}
 The positive definite kernel $\mathcal{K}(x,y)$ corresponds to an inner product 
$\langle  \phi_{x},\phi_{y}\rangle_{\mathcal{H}}$ in some feature space $\mathcal{H}$. 
The class of conditionally positive definite (cpd) kernel is less well known. Nevertheless 
this class is of great interest since SCh\"{o}lkopf show in (P. J. Moreno, P. P. Hu, and N Vasconcelos 2003) that all translation invariant kernel methods 
can also use the larger class of cpd kernels. Therefore we give a short summary of this type of kernels and their
connection to Hilbertian metrics.\\

\begin{defn}
A real valued function $\mathcal{K}$ on $\mathcal{X}\times\mathcal{X}$ is positive definite (pd) (resp. conditionally positive definite (cpd) 
if and only if $\mathcal{K}$ is symmetric and $\sum_{i,j}^{n}c_{i}c_{j}\mathcal{K}(x_{i},x_{j})\geq0$,for all $n\in \mathbb{N},x_{i}\in \mathcal{X},i=1,...,n$,
and for all $c_{i}\in \mathbb{R},i=1,...,n,$ (resp. for all $c_{i}\in \mathbb{R},i=1,...,n$ with $\sum_{i}^{n}c_{i}=0$).
\end{defn}

The following theorem describes the class of Hilbertian metrics:

\begin{thm}  (I. J Schoenberg 1938)\\
A metric space $(\mathcal{X},d)$ can be embedded isometrically into a Hilbert space if and only if 
$-d^2(x,y)$ is conditionally positive definite (cpd).
\end{thm}

\begin{lem}\label{lem:are}
(J. P. R. Christensen, C. berg and P. Ressel 1984) \\
Let $\mathcal{K}$ be a kernel defined as $\mathcal{K}(x,y)=\widehat{k}(x,y)-\widehat{k}(x,x_{0})-\widehat{k}(x_{0},y)+\widehat{k}(x_{0},x_{0})$, where $x_{0}\in \mathcal{X}$.
Then $\mathcal{K}$ is pd if and only if $\widehat{k}$ is cpd.
\end{lem}

Similar to pd kernel one can also characterize cpd kernel. Presently one can write all cpd kernel in the form :
$\mathcal{K}(x,y)=\displaystyle\frac{-1}{2}||\phi_{x}-\phi_{y}||_{\mathcal{H}}^{2}+f(x)+f(y)$. The cpd kernel corresponding to Hilbertian (semi)-metrics 
are characterized by $f(x)=0$ for all $x \in \mathcal{X}$, whereas if $\mathcal{K}$ is pd it follows that $f(x)=\displaystyle\frac{1}{2}\mathcal{K}(x,x)\geq 0$.
We also would like to point out that for SVM the class of Hilbertian (semi)-metrics is more important than the class of pd kernels.
Namely one can show,  (see M. Hein and O. Bousquet 2003),  which the solution and optimization problem of the SVM only depends on the Hilbertian (semi)-metric,  which is implicitly defined   by each pd kernel. Moreover a whole family of pd kernels induces the same metric. 
 Hilbertian metrics since, using \textbf{lemma \ref{lem:are}}, one can always define a corresponding pd kernel.
 Nevertheless for the convenience of the reader we will often explicity state the corresponding pd kernels.

\subsection{Hilbertian Metrics on Probabilitity Measures}

For simplicity, we are dealing with the case of discrete probability measures
 on $D=\{1,2,...,N\}$ ,where $1\leq N \leq \infty$. Given a Hilbertian metrics d on $\mathbb{R}_{+}$ it is easy 
to see that the metric on $d_{\mathcal{M}_{+}^{1}}$ given by $D_{\mathcal{M}_{+}^{1}}^{2}(P,Q)=\sum_{i=1}^{N}d_{\mathbb{R}_{+}}^{2}(p_{i},q_{i})$
is a Hilbertian metric on $\mathcal{M}_{+}^{1}(D)$. The following proposition extends the simple discrete case to the general case of 
a Hilbertian metrics on a probability space 
$\mathcal{X}$. In order to simplify the notation we define $p(x)$ to be the Radon-Nikodym 
derivative $(dP/d\mu)(x)$ of P with respect to the dominating measure $\mu$.
\begin{prop}
Let P and Q be two probability measures on $\mathcal{X}$, $\mu$ an arbitrary dominating measure of P and Q 
and $d_{\mathbb{R}_{+}}$ (where $d_{\mathbb{R}_{+}}$ is the positive part of the real line with 0 included) a $\displaystyle1/2$-homogeneous Hilbertian metrics on $\mathbb{R}_{+}$. Then $D_{\mathcal{M}_{+}^{1}(\mathcal{X})}^{2}$ defined as 
\begin{equation}
 D_{\mathcal{M}_{+}^{1}}^{2}(P,Q):=\int_{\mathcal{X}}d_{\mathbb{R}_{+}}^{2}(p(x),q(x))d_{\mu}(x)
\end{equation}
is a Hilbertian metrics on $\mathcal{M}_{+}^{1}(\mathcal{X})$. $D_{\mathcal{M}_{+}^{1}(\mathcal{X})}^{2}$ is independent of the dominating measure $\mu$.
\end{prop}
\begin{pro}First we show by using the $1/2$-homogeneity of $d_{\mathbb{R}_{+}^{1}}$ is independent  of the dominating measure $\mu$. We have 
$$\int_{\mathcal{X}}d_{\mathcal{M}_{+}^{1}}^{2}(\frac{dP}{d\mu},\frac{dQ}{d\mu})d\mu=\int_{\mathcal{X}}d_{\mathcal{M}_{+}^{1}}^{2}(\frac{dP}{d\nu}\frac{d\nu}{d\mu},
\frac{dQ}{d\nu}\frac{d\nu}{d\mu})\frac{d\mu}{d\nu}d\nu=\int_{\mathcal{X}}d_{\mathcal{M}_{+}^{1}}^{2}(\frac{dP}{d\nu},\frac{dQ}{d\nu})d\nu$$
\end{pro}
where we use that $d_{\mathbb{R}_{+}}^{2}$ is $1$-homogeneous. It is easy that $-D_{\mathcal{M}_{+}^{1}}^{2}(\mathcal{X})$ is conditionally positive definite, 
simplicity take for every $n \in \mathbb{N},P_{1},....,P_{n}$
 the dominating measure $\frac{\sum_{i=1}^{n}P_{i}}{n}$ and use that $-D_{\mathbb{R}_{+}}^{2}$ is conditionally positive definite.
 
\begin{rem} 
It is in principe very easy to build hilbertian metrics on $\mathcal{M}_{+}^{1}(\mathcal{X})$ using  arbitrary Hilbertian metrics
 on $\mathbb{R}_{+}$ and plugging it into the \textbf{definition 1}.  
\end{rem}
 But the key property of the method we propose is the independence of the metric $d$ on $\mathcal{M}_{+}^{1}(\mathcal{X})$ of 
 the dominating measure. That is we have generated a metric which is invariant with respect to general coordinate transformations on $\mathcal{X}$, 
 therefore we call it a covariant metric.

\subsection{$\lambda$-homogeneous Hilbertian Metrics and Positive Definite Kernels on $\mathbb{R}_{+}$}

In this paper we consider the class of Hilbertian metrics on probability measure, therefore the Hilbertian metrics  on $\mathbb{R}_{+}$ is   the  main element of our approche. 
 This is  the event we requiere that the Hilbertian metrics on $\mathbb{R}_{+}$ is $\lambda$-homogeneous.
 The class of $\lambda$-homogeneous Hilbertian metrics on $\mathbb{R}_{+}$ was characterized by Fuglede:
  \begin{defn}(Tops$\phi$e 2003) and B.Fuglede 2004)\\
  A Hilbertian metrics si $\lambda$-homogeneous if and only if $d^{2}(cp,cq)=c^{\lambda}d^{2}(p,q)$ for all $c\in \mathbb{R}_{+}$ .
  \end{defn}
 
 \begin{thm}(B. Fuglede 2004)\\
 A symmetric function $d : \mathbb{R}_{+}\times \mathbb{R}_{+}\rightarrow \mathbb{R}_{+}$ with $d(x,y)=0 \Longleftrightarrow x=y$ is a $\gamma$-homogeneous, continuous Hilbertian metrics d on $\mathbb{R}_{+}$ 
 if and only if there existe a (necessarily unique) non-zero bounded measure $\rho \geq 0$ on $\mathbb{R}_{+}$ such that $d^2$ can be written as 
 \begin{equation}
  d^2(x,y)=\int_{\mathbb{R}_{+}}|x^{(\gamma+i\lambda)}-y^{\gamma+i\lambda}|^{2}d\rho(\lambda)
 \end{equation}
 \end{thm}

 Using \textbf{lemma \ref{lem:are} }we define the corresponding class of pd kernel on $\mathbb{R}_{+}$ by choosing $x_{0}=0$. We will see later that this corresponds to choosing the zero-measure as origin of the RKHS (reproducing kernel Hilbert space).
 
  \begin{cor}
 A symmetric function $k :  \mathbb{R}_{+}\times \mathbb{R}_{+}\rightarrow \mathbb{R}_{+}$ with $k(x,y)=0\Longleftrightarrow x=0$ is a $2\gamma$-homogeneous continuous pd kernel $\mathcal{K}$ on $\mathbb{R}_{+}$ 
 if and only if there exists a (necessarily unique ) non-zero bounded symmetric measure $\kappa\geq0$ on $\mathbb{R}$ such that $\mathcal{K}$ is given as 
 \begin{equation}
  \mathcal{K}(x,y)=\int_{\mathbb{R}}x^{(\gamma+i\lambda)}y^{(\gamma-i\lambda)}d\kappa(\lambda)
 \end{equation}
 
 \end{cor}
 \begin{pro}
 If we have the form given in (2), then it is obviously $2\gamma$-homogeneous and since $\mathcal{K}(x,y)=x^{2\gamma}\kappa(\mathbb{R})$ we have have $\mathcal{K}(x,x)=0\Longleftrightarrow x=0$. 
 
 The other direction follows by first noting 
 that $\mathcal{K}(0,0)=\langle  \phi_{0},\phi_{0}\rangle=0$ and then by appying theorem 2, where $\kappa$ is the symmetrized version of $\rho$ around the origin, together with \textbf{lemma \ref{lem:are} }
 and $\mathcal{K}(x,y)=\langle \phi_{x},\phi_{y}\rangle=\frac{1}{2}(-d^{2}(x,y)+d^{2}(x,0)+d^{2}(y,0)).$    $\Box$
 \end{pro}

   Andrezej Cichocki, Sergio Cruees and Shun-ichi Amari (January 2011) proposed an interesting two-parameter family of metrics, the AB-divergence defined: 
     \begin{defn}
The function $d:\mathbb{R}_{+}\times \mathbb{R}_{+}\rightarrow \mathbb{R}$ defined as:
\begin{equation}
 d_{AB}^{(\alpha,\beta)}(x,y)=\left\{
  \begin{array}{cccc}
  && -\frac{1}{\alpha\beta}(x^{\alpha}y^{\beta}-\frac{\alpha}{(\alpha+\beta)}
  x^{\alpha+\beta}-\frac{\beta}{(\alpha+\beta)}y^{\alpha+\beta}) \hspace{0.2cm}for\hspace{0.2cm} \alpha,\beta,\alpha+\beta\neq 0\cr\\
  &&\frac{1}{\alpha^{2}}(x^{\alpha}\log (\frac{x^{\alpha}}{y^{\alpha}})-x^{\alpha}+y^{\alpha})\hspace{0.5cm}for\hspace{0.5cm} \alpha \neq 0, \beta=0 \cr\\
  && \frac{1}{\alpha^{2}}(\log( \frac{y^{\alpha}}{x^{\alpha}})+(\frac{y^{\alpha}}{x^{\alpha}})^{-1}-1)\hspace{0.5cm}for \hspace{0.5cm}\alpha=-\beta\neq0\cr\\
  && \frac{1}{\beta^{2}}(y^{\beta}\log (\frac{y^{\beta}}{x^{\alpha}})-y^{\beta}+x^{\alpha}) \hspace{0.5cm}for\hspace{0.5cm}\alpha=0;\beta\neq0 \cr\\
  && \frac{1}{2}(\log x -\log y)^{2}\hspace{0.5cm}for\hspace{0.5cm}\alpha,\beta=0 \cr 
 \end{array}
 \right.
\end{equation}
where $ d_{AB}^{(\alpha,\beta)}(x,y)$ is a divergence on $\mathbb{R}_{+}$
\end{defn}
 
   \subsection{A brief recall of  Support Vector Machine (SVM)}
  SVM were developed by Cortes and Vapnik (1995) for binary classification. Their approach may be roughly sketched as follows:
\begin{itemize}
\item class separation: basically, we are looking for the optimal separating hyperplane between the two classes by maximizing the margin between the classes closest points, the points lying on the boundaries are called support vectors, and the middle of the margin is our optimal separating hyperplane;
\item nonlinearity: when we cannot find a linear separator, data points are projected into an (usually) higher-dimensional space where the data points effectively become linearly separable (this  projection is relised via kernel techniques);
\item problem solution: the whole task can be formulated as a quadratic optimization problem which can be solved by known techniques etc.
\end{itemize}
  An implicit mapping $\Phi$ was used by SVM of the input data into a high-dimensional fearture space defined by a kernel function, the inner product $\langle \Phi (x),\Phi (y) \rangle$ between the images of two data points $x,y$ was returing by the function in the feature space. The kernel function can be represented as 
  \begin{equation}
  \mathcal{K}(x,y)=\langle \Phi (x),\Phi (y) \rangle
\end{equation}   
where $\Phi : X\rightarrow H $ is the projection function, this function project x and y into the feature space H.

Relationship between the kernel method and SVM: Sch\"{o}lkopf showed that the class 
of cpd kernel can be used in SVM due to the translation invariant of the maximal margin problem in the RKHS, and the kernel can be used in SVM to the classification, the regression etc if we found a good kernel function . 
The advantage of kernel method and SVM is that we can found and used a kernel for a problem particular that could be applied directly to data without the need for a feature extraction process.
 This was used in (M. Hein and O. Bousquet 2003) to show that the properties of the SVM only depend on the Hilbertian metrics. That is all cpd kernel are 
generated by a Hilbertian metric $d(x,y)$ through $\mathcal{K}(x,y)=-d^2(x,y)+g(x)+g(y)$ where $g: \mathcal{X} \longrightarrow \mathbb{R}$ and the solution of the SVM only depends on the Hilbertian metric $d(x,y)$.

\section{Main results}

Generally the AB-metrics is not symmetric. We extend and improve this family an two-parameter symmetric.
 The metrics we propose is very interesting since it is a 
symmetric and smoothed variant from AB-metrics. This allows us to recover all proprety in Hilbertian
metrics on $\mathcal{M}_{+}^{1}(\mathcal{X})$  from  the family of two-parameter.
 \begin{thm}
  The function $d:\mathbb{R}_{+}\times\mathbb{R}_{+}\rightarrow \mathbb{R}_{+}$ defined as :
  \begin{equation}
   d_{ABS}^{\alpha,\beta}(x,y)=\left\{
   \begin{array}{cccc}
   &&\frac{1}{\alpha\beta}(x^{\alpha}-y^{\alpha})(x^{\beta}-y^{\beta})\hspace{0.2cm} for \hspace{0.2cm}\alpha,\beta\neq0\cr\\
   &&\frac{1}{\alpha^{2}}(x^{\alpha}-y^{\alpha})\log(\frac{x^{\alpha}}{y^{\alpha}}) \hspace{0.2cm}for \hspace{0.2cm}\alpha\neq 0,\beta=0\cr\\
   &&\frac{1}{\alpha^{2}}((x^{\alpha}-y^{\alpha})\log(\frac{x^{\alpha}}{y^{\alpha}})+(\frac{x^{\alpha}}{y^{\alpha}})^{-1}+
   (\frac{y^{\alpha}}{x^{\alpha}})^{-1}-2)\hspace{0.1cm}for\hspace{0.1cm} \alpha=-\beta\neq 0\cr\\
   &&\frac{1}{\beta^{2}}(x^{\beta}-y^{\beta})\log(\frac{x^{\beta}}{y^{\beta}})\hspace{0.2cm}for\hspace{0.2cm}\alpha=0,\beta\neq 0\cr\\
   &&\frac{1}{2}(\log x-\log y)\hspace{0.2cm}for \hspace{0.2cm}\alpha=\beta=0\cr\\   
   \end{array}
   \right.
  \end{equation}
is a $\gamma$-homogeneous Hilbertian metrics on $\mathbb{R}_{+}$,note that $d_{ABS}^{(\alpha,\beta)}$ is symmetric.
\end{thm}
\begin{pro}
The proof for the symmetry is trivial because this function is symmetric by construction and $d_{ABS}^{(\alpha,\beta)}(cx,cy)=c^{\alpha +\beta}d_{ABS}^{(\alpha,\beta)}(x,y)$ where $\gamma=\alpha + \beta $ then $d_{ABS}^{(\alpha ,\beta)}$ is $\gamma$-homogeneous. Second for simplicity note that $\mathcal{K}(x,y)=-d^{2}(x,y)$, where $d^{2}$ is a Hilbertian metrics. The all conditions for theorem of Schoenberg satisfied we have $-d^{2}$ cpd.$\square$ 
\end{pro}

We can now apply the principle to construct Hilbertian metrics on
 $\mathcal{M}_{+}^{1}(\mathcal{X})$, of building  Hilbertian metrics on $\mathcal{M}_{+}^{1}(\mathcal{X})$ and use the
family of $\gamma$-homogeneous Hilbertian metrics $d_{ABS}^{(\alpha,\beta)}$ on $\mathbb{R}_{+}$.

\begin{defn}

 We proposed two ways to build the symmetry divergence:\\ Thype-1 :
 \begin{equation}
  D_{ABS}^{(\alpha,\beta)}=\frac{1}{2}[D_{AB}^{(\alpha,\beta)}(P,Q)+D_{AB}^{(\alpha,\beta)}(Q,P)]
 \end{equation}
 Type-2:
 \begin{equation}
  D_{ABS}^{(\alpha,\beta)}(P,Q)=\frac{1}{2}[D_{AB}^{(\alpha,\beta)}(P,\frac{P+Q}{2})+D_{AB}^{(\alpha,\beta)}(Q,\frac{P+Q}{2})]
 \end{equation}
 \end{defn}
 For the construction ABS-divergence we used the definition above and we apply the \textbf{proposition} (1) from the section(2). We obtain 
 the symmetric ABS-divergence (Type-1) defined as:
 \begin{defn}: Let P and Q two probability measures on $\mathcal{X}$ (probability space) and $\mu$ an dominating measure of P, Q and $d_{ABS}^{(\alpha,\beta)}$ a $\gamma$-homogeneous Hilbertian metrics on $\mathbb{R}_{+}$, then $D_{ABS}^{(\alpha,\beta)}$ defined as:
 \begin{equation}
   D_{ABS}^{(\alpha,\beta)}(P,Q)=\left\{
  \begin{array}{cccc}
   && \frac{1}{\alpha\beta}\int_{\mathcal{X}} (p^{\alpha}-q^{\alpha})(p^{\beta}-q^{\beta})d\mu(x)\hspace{0.2cm}for
   \hspace{0.2cm}\alpha\neq0,\beta\neq0 \cr\\
   && \frac{1}{\alpha^{2}}\int_{\mathcal{X}} (p^{\alpha}-q^{\alpha})\log(\frac{p^{\alpha}}{q^{\alpha}})d\mu(x)
   \hspace{0.2cm}for \hspace{0.2cm}\alpha\neq 0,\beta=0\cr\\
   && \frac{1}{\alpha}\int_{\mathcal{X}} ((p^{\alpha}-q^{\alpha})\log(\frac{p^{\alpha}}{q^{\alpha}})+
   \frac{q^{\alpha}}{p^{\alpha}}+\frac{p^{\alpha}}{q^{\alpha}}-2)d\mu(x) \hspace{0.2cm}for \hspace{0.2cm}\alpha=-\beta\neq 0\cr\\
   && \frac{1}{\beta^{2}} \int_{\mathcal{X}} (p^{\beta}-q^{\beta})\log(\frac{p^{\beta}}{q^{\beta}})d\mu(x)\hspace{0.2cm}for \hspace{0.2cm}\alpha=0,\beta\neq0\cr\\
   && \frac{1}{2}\int_{\mathcal{X}} (\log x-\log y)d\mu(x) \hspace{0.2cm}for \hspace{0.2cm} \alpha=\beta=0\cr\\
  \end{array}
     \right.
 \end{equation}
 is a $\gamma$-homogeneous Hilbertian metrics on $\mathcal{M}_{+}^{1}(\mathcal{X})$.
 \end{defn}

 The ABS-divergence has the following basic properties:
 \begin{propr}:\\  
 \begin{enumerate}
  \item Convexity : $D_{ABS}^{(\alpha,\beta)}(P,Q)$ is convex with respect to both P and Q.
  \item Strict Positvity: $D_{ABS}^{(\alpha,\beta)}(P,Q)\geq 0$ and $D_{ABS}^{(\alpha,\beta)}(P,Q)=0$ if and only if $P=Q$.
  \item Continuity: The ABS-divergences is continuous function of real variant $(\alpha,\beta)$ in the whole range 
  including singularities.
  \item Symmetric: $D_{ABS}^{(\alpha,\beta)}(P,Q)=D_{ABS}^{(\alpha,\beta)}(Q,P)$
   \item $\gamma$-homogeneous: $D_{ABS}^{(\alpha,\beta)}(cP,cQ)=c^{\alpha+\beta}D_{ABS}^{(\alpha,\beta)}(P,Q)$ 
 \end{enumerate}
 \end{propr}

We used the instrument of building  Hilbertian metrics on $\mathcal{M}_{+}^{1}(\mathcal{X})$  
and use the family of ($\alpha+\beta$)-homogeneous Hilbertian metrics $d_{ABS}^{(\alpha,\beta)}$ on $\mathbb{R}_{+}$. 
This yield as special case the following measures on $\mathcal{M}_{+}^{1}(\mathcal{X})$.

$$ D_{ABS}^{\alpha,\beta}(P,Q)=\int_{\mathcal{X}} \varphi_{\alpha ,\beta}(p(x)q(x))d\mu (x)$$

\begin{center}
\begin{tabular}{|c|c|c|}
\hline
Divergence $D_{ABS}^{\alpha,\beta}(P,Q)$& function $\varphi_{\alpha ,\beta}(p(x)q(x))$& Name\\
& & \\
\hline
$D_{ABS}^{(1,1)}(P,Q)$ & $(p(x)-q(x))^{2}$ & Euclidian\\
& & \\
\hline
$D_{ABS}^{(1/2,1)}(P,Q)$  & $2(\sqrt{p(x)}-\sqrt{q(x)})(p(x)-q(x))$& $V_{1}$-Hellinger \\
& & \\
\hline
$D_{ABS}^{(1/2,-1)}(P,Q)$  & $2(\sqrt{p(x)}-\sqrt{q(x)})\frac{(p(x)-q(x))}{p(x)q(x)}$ & $V_{2}$-Hellingher\\
& & \\
\hline
$D_{ABS}^{(1/2,1/2)}(P,Q)$ &$4(\sqrt{p(x)}-\sqrt{q(x)})^{2}$ & Hellinger\\
& & \\
\hline
$D_{ABS}^{(1,0)}(P,Q)$ &$(p(x)-q(x))\log(\frac{p(x)}{q(x)})$ & Jeffrey\\
& & \\
\hline

\end{tabular}\\
\title{Table (a): $D_{ABS}^{\alpha,\beta}(P,Q)$ divergence}
\end{center}

 $D_{ABS}^{(1,1)}$ corresponds to the square of euclidian metric, $D_{ABS}^{(1/2,1/2)}$ corresponds to the Hellinger metric which
 is well known in the statistics community, $D_{ABS}^{(1,0)}$ correspond to the Jeffreys metric, $ D_{ABS}^{(1/2,1)}$ $V_{1}$-Hellinger and $D_{ABS}^{(1/2,-1)}$  $V_{2}$-Hellinger is a variant of Hellinger metrics.\\
 For completeness we also give the corresponding pd kernels on $\mathcal{M}_{+}^{1}(\mathcal{X})$, 
where we take in \textbf{lemma \ref{lem:are}} the zero measure as $x_{0}\in \mathcal{M}_{+}^{1}(\mathcal{X})$. 
This choice seems strange at first since we are dealing with probability measures. 
But in fact the whole framework presented in this paper can easily be extended to all finite, positive 
measure on $\mathcal{X}$. For this, set zero measure is a natural choice of the origin.
\[ \mathcal{K}_{(1,1)}(P,Q)=\int_{\mathcal{X}} p(x)q(x)d\mu(x)\]
\[\mathcal{K}_{(1/2,1)}(P,Q)=\int_{\mathcal{X}} (q(x)\sqrt{p(x)}+p(x)\sqrt{q(x)})d\mu(x)\]
\[ \mathcal{K}_{(1/2,1/2)}(P,Q)=4\int_{\mathcal{X}} \sqrt{p(x)q(x)}d\mu(x)\]
\[ \mathcal{K}_{(1,0)}(P,Q)=\frac{1}{2}( \int_{\mathcal{X}} (q(x)-p(x))\log(\frac{p(x)}{q(x)})-p(x)-q(x))d\mu(x)\]

Using two-parameters is so difficult in practice that is the reason why 
we will proposed a one-parameter family to improve the ABS-divergence.

 \begin{prop}
  The function $d :\mathbb{R}_{+}\times\mathbb{R}_{+}\longrightarrow \mathbb{R}_{+}$ defined as :
  \begin{equation}
   d_{t}^{2}(x,y)=\left\{
   \begin{array}{cccc}  
 && \frac{1}{2}(\frac{x^{t}-y^{t}}{t})^2 \hspace{0.5cm} for\hspace{0.2cm}t\neq0 \cr\\
 && \frac{1}{2}(\log x-\log y) \hspace{0.5cm}for\hspace{0.2cm}t=0\cr
 \end{array}
     \right.
   \end{equation}
is a $2t$-homogeneous Hilbertian metric on $\mathbb{R}_{+}$ if $t\neq 0$. 
\end{prop}

\begin{pro}
Note that $d_{t}^{2}$ is symmetric by construction and 
it's easy verified the property of Hilbertian metric. Therfore we show that: $d_{t}^{2}(cx,cy)=\displaystyle\frac{1}{2}(\frac{(cx)^{t}-(cy)^{t}}{t})^2=\displaystyle\frac{c^{2t}}{2}(\frac{x^{t}-y^{t}}{t})^2=c^{2t}d_{t}^{2}(x,y)$.$\square$
\end{pro} 
We used the proposition 1 of building Hilbertian metrics on $\mathcal{M}_{+}^{1}(\mathcal{X})$  
and use the family of $2t$-homogeneous Hilbertian metrics $d_{t}^{2}$ on $\mathbb{R}_{+}$. 
Therefor we obtain as special case the following measures on $\mathcal{M}_{+}^{1}(\mathcal{X})$.
\[ D_{t}^{2}(P,Q)=\int_{\mathcal{X}} \varphi_{t}(p(x)q(x))d\mu (x)\]

\begin{center}

\begin{tabular}{|c|c|c|}
\hline
Divergence $D_{t}^{2}(P,Q)$& function $\varphi_{t}(p(x)q(x))$& Name\\
& & \\
\hline
$D_{1}^{2}(P,Q)$ & $\frac{1}{2}(p(x)-q(x))^{2}$ & Euclidian\\
& & \\
\hline
$D_{-1}^{2}(P,Q)$  & $\frac{1}{2} (\frac{1}{p(x)}-\frac{1}{q(x)})^{2}$ & S-Euclidian \\
& & \\
\hline
$D_{-\frac{1}{2}}^{2}(P,Q)$  & $2(\sqrt{p(x)}-\sqrt{q(x)})^{2}$ &  Hellinger\\
& & \\
\hline
$D_{\frac{1}{2}}^{2}(P,Q)$ &$2(\frac{1}{\sqrt{ p(x)}}-\frac{1}{\sqrt{q(x)}})^{2}$ & S-Itakura Saito\\
& & \\
\hline

\end{tabular}\\
\title{Table (b): Divergence using $2t$-homogeneous Hilbertian metric}
\end{center}

$D_{1}^{2}$ correspond to the square of euclidian metric, $D_{-1}^{2}$ is an other version of euclidian metric,
$D_{1/2}^{2}$ corresponds 
two Hellinger mertric. The Hellinger metric is well known in the statistics community.
$D_{-\frac{1}{2}}$ is a symmetrized Itakura-Saito distance (called also the COSH distance) modified.

For completeness we also give the corresponding pd kernels on $\mathcal{M}_{+}^{1}(\mathcal{X})$, 
where we take in \textbf{lemma \ref{lem:are}} the zero measure as $x_{0}\in \mathcal{M}_{+}^{1}(\mathcal{X})$. 
This choice seems strange at firts since we are dealing with probability measures. 
But in fact the whole framework presented in this paper can easily be extended to all finite, positive 
measures on $\mathcal{X}$. For this set zero measure is a natural choice  the origin.
\[\mathcal{K}_{t}(P,Q)=\frac{1}{2}\int_{\mathcal{X}}\frac{x^ty^t}{t^2}d\mu \]
\[ \mathcal{K}_{1}(P,Q)=\frac{1}{2}\int_{\mathcal{X}}p(x)q(x)d\mu(x)\]
\[\mathcal{K}_{-1}(P,Q)=\frac{1}{2}\int_{\mathcal{X}}\frac{1}{p(x)q(x)}d\mu(x)\]
\[\mathcal{K}_{\frac{1}{2}}(P,Q)=2\int_{\mathcal{X}}\sqrt{p(x)q(x)}d\mu(x)\]
\[\mathcal{K}_{-\frac{1}{2}}(P,Q)=2\int_{\mathcal{X}}\frac{1}{\sqrt{p(x)q(x)}}d\mu(x)\]

\section{Numerical studies}
To show the interest of our study the examples of applications have been proposed. For the SVM we made studies on the classification of the genes and on the sex of the cats knowing the weight of the heart and the body. As regards classification of images, we use our divergencs to separate them into two classes.
\subsection{Application in SVM}
The performance of metrics and kernels has been compared in classification using some  data sets.
 All data sets were split into a training (80\%) and a test (20\%)   set. For the problem we use SVM method. For all experiments we use the one-parameter family $d^{2}_{t}$ of hilbertian metric, the  corresponding kernel cpd is $\mathcal{K}=-D^{2}_{t}$ with varying penalety constants $C$ in the SVM, and we use the transformed kernel (gaussian transformation):
 \[\mathcal{K}_{t}(P,Q)=\displaystyle e^{-D^{2}_{t}(P,Q)/2 \sigma^{2}}\]
The test error was evaluated by the best parameters $C$ and $\sigma$. The best constant penality $C$ and $\sigma$  was found by cross-validation.

We evaluated the performance of the proposed metrics and kernels in three classification tasks. Firstly we generated a artificiel data and we consider the test error for kernels proposed. Secondly as a real-world application, let us test the ability of SVM to predict the class of a tumour from gene expression data. We use a publicly available datasets of gene expression data for 128 different individuals with acute lymphoblastic leukemia (ALL). Here we focus on predicting the type of the disease (B-cell or T-cell). Therfore we test a SVM classifier for cancer diagnosis from gene expression data, and we test the ability of a SVM to predict the class of the disease from gene expression. Finally we apply the data from support functions and datasets for Venables and Ripley's MASS. We use the anatomical data from domestic cats, the heart and body weights of samples of male and famale cats used for digitalis experiments. The cats were all adult, over 2 kg body weight. We presented the classification error according to the sex of the cats.

 The tables shows the test errors for the kernels corresponding to $t=\{-1;-1/2;1/2;1\}$ from the ABS-divergence resp. and their gaussian transformation. The first line shows the kernels directly (dir) and the second line the gaussian transformation (tran).

\begin{center}
\begin{tabular}{|c||c|c|c|c|}
\hline
Divergence & Euclidian  & Hellinger & Itakura-Saito & S-Euclidian \\
 & error \hspace{0.2cm} C \hspace{0.2cm}$\sigma$& error  \hspace{0.2cm} C \hspace{0.2cm} $\sigma$ & error  \hspace{0.2cm} C  \hspace{0.2cm} $\sigma$ & error  \hspace{0.2cm} C  \hspace{0.2cm} $\sigma$ \\

\hline
Data \hspace{0.15cm} (dir)&0.0001\hspace{0.2cm} 10\hspace{0.2cm} - & 0.005\hspace{0.2cm} 10\hspace{0.2cm} -- & 0.26\hspace{0.2cm}100\hspace{0.2cm} --  & 0.13\hspace{0.2cm} 10\hspace{0.2cm} - \\
Artificial\hspace{0.15cm}(tran)&0.0001\hspace{0.2cm} 10\hspace{0.2cm} 1.5 & 0.005\hspace{0.2cm} 10\hspace{0.2cm} 0.5 & 0.125\hspace{0.2cm} 100\hspace{0.2cm} 1.5 & 0.17\hspace{0.2cm} 100\hspace{0.2cm} 1.5 \\ 
\hline
\end{tabular}\\
\title{Table 1:   Test error using data artificial}
\end{center}
 
Table 1 shows that the errors committed using the conditionally defined positive $\mathcal{K}$ and the kernel transformed $\mathcal{K}_t$ are in the same proportion for the Euclidean divergence and the Hellinger divergence with $\sigma= 1.5$ and the constant C = 10 .  While for the Itakura-saito divergence the errors committed in the classifications with the $\mathcal{K}_t$ kernel are smaller than that of the $\mathcal{K}$ kernel with $\sigma = 1.5$ and C = 100. So the classification with the transformed core is better than the one used directly. For the S-Euclidean divergence by varying the constant C, C = 10 for the kernel $\mathcal{K}$ and C = 100 for the kernel transforms $\mathcal{K}_t$ , we find that the transformed nucleus offers a better classification result. Thus in all cases we notice that the transformed nuclei give the best results. For the sake of using the transformed nuclei for artificial data.

\begin{center}
\begin{tabular}{|c||c|c|c|c|}
\hline
Divergence & Euclidian  & Hellinger & Itakura-Saito & S-Euclidian \\
 & error \hspace{0.2cm} C \hspace{0.2cm}$\sigma$& error  \hspace{0.2cm} C \hspace{0.2cm} $\sigma$ & error  \hspace{0.2cm} C  \hspace{0.2cm} $\sigma$ & error  \hspace{0.2cm} C  \hspace{0.2cm} $\sigma$ \\
 \hline
Data"ALL"\hspace{0.1cm}(dir)&0.679 \hspace{0.2cm} 10 \hspace{0.2cm}-& 0.414\hspace{0.2cm} 10 \hspace{0.2cm} - & 0.257 \hspace{0.2cm} 10 \hspace{0.2cm} - & 0.461 \hspace{0.2cm} 10 \hspace{0.2cm} - \\
gene \hspace{0.4cm}(tran)&0.0234\hspace{0.2cm} 10 \hspace{0.2cm}0.5 & 0.156\hspace{0.2cm} 100 \hspace{0.2cm} 1.5 & 0.0468 \hspace{0.2cm} 100 \hspace{0.2cm} 1.5 & 0.164 \hspace{0.2cm} 100\hspace{0.2cm} 1.5 \\
\hline 
\end{tabular}\\
\title{Table 2: test error using data "ALL" gene}
\end{center} 
With the data gene (ALL), we find that the kernel transformed offers the best resutats for the classification. However with the use of kernels $\mathcal{K}$ it is the kernel constructed with the divergence of Itakoura-Saito that gives the best classification results, followed by Hellinger, S-Euclidean and the Euclidean divergence. Whereas if we use kernel transformed, it is those built with Euclidean divergence that give the best results, followed by that of Itakura-Saito, Hellinger and S-Euclidian.
\begin{center}
\begin{tabular}{|c||c|c|c|c|}
\hline
Divergence & Euclidian  & Hellinger & Itakura-Saito & S-Euclidian \\
 & error \hspace{0.2cm} C \hspace{0.2cm}$\sigma$& error  \hspace{0.2cm} C \hspace{0.2cm} $\sigma$ & error  \hspace{0.2cm} C  \hspace{0.2cm} $\sigma$ & error  \hspace{0.2cm} C  \hspace{0.2cm} $\sigma$ \\
\hline
Data"MASS"\hspace{0.15cm}(dir)&0.236 \hspace{0.2cm} 1 \hspace{0.2cm}-& 0.326\hspace{0.2cm} 1 \hspace{0.2cm} - & 0.340 \hspace{0.2cm} 1 \hspace{0.2cm} - & 0.326 \hspace{0.2cm} 10 \hspace{0.2cm} - \\
cats\hspace{0.3cm}(tran) &0.194\hspace{0.2cm} 1 \hspace{0.2cm}0.5 & 0.326\hspace{0.2cm} 10 \hspace{0.2cm} 0.5 & 0.368 \hspace{0.2cm} 1 \hspace{0.2cm} 0.5 & 0.181 \hspace{0.2cm} 1 \hspace{0.2cm} 0.5 \\
\hline
\end{tabular}\\
\title{Table 3: Test error using data cats "MASS"}
\end{center}

With the data cats (MASS), if we use kernel $\mathcal{K}$ we have almost the same results for all measures of divergence. However with the modified kernel the classification obtained with the S-Euclidean divergence and the Euclidean divergence offer the best results. It should also be noted that for the divergence of Itakura-Saito it is the kernel $\mathcal{K}$ that gives the best results.
In conclusion we retain that the choice of the best method depends on data and constants.

\subsection{Applications in imagess classifications}

In this work we apply the above proposed algorithm  and also implement with the metrics proposed in the section above, to color image segmentation. The divergence we proposed gives a good classification with different threshold (k). In our experiments we used the data from the image (a) and apply our algorithm with different threshold. From these figures, we can see the results experimented by our algorithm. These results have been  obtained with our divergences following some (k) values.\\

\vspace{1cm}
\begin{center}
\begin{tabular}{ccc}
\hline
\hline 
&\textbf{Algorithm}&
\\\hline
\hline
$X \in M_{l,c}(\mathbb{R})$& &matrix origin
\\
$X^{'} \in M_{l,c}(\mathbb{R})$ && matrix sortie
\\
$X^{'}$&$\longleftarrow$& matrix null order $l\times c$
\\
for i & rang($1,l-1$):&
\\
$P_{1}$&$\longleftarrow$& $X_{i,j-1}$
\\
$P_{2}$&$\longleftarrow$&$X_{i-1,j}$
\\
If &norm$(P_{1},P_{2})<k$&
\\
$P_{0}$&  $\longleftarrow$& $(225,225,225)$
\\
else:&&
\\
$P_{0}$& $\longleftarrow$ &$(0,0,0)$
\\
 $X^{'}$&$\longleftarrow$& $P_{0}$
\\\hline
\hline
\end{tabular}

\end{center}

\subsubsection{Facial image segmentation}
 \begin{center}
\begin{tabular}{ccc}

\includegraphics[width=3cm]{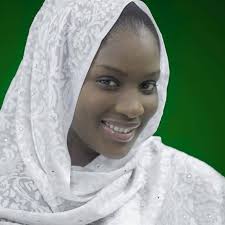}\title{a}&\includegraphics[width=3cm]{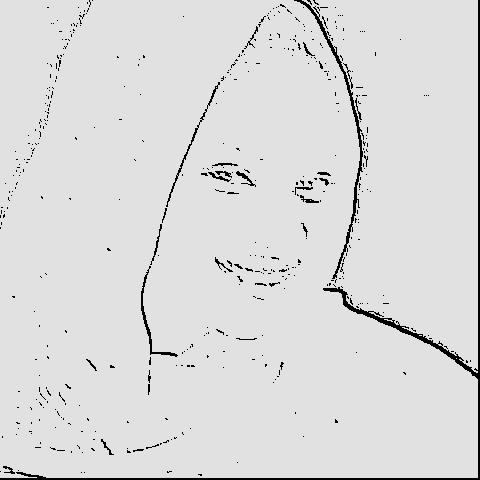}\title{b}\\
\includegraphics[width=3cm]{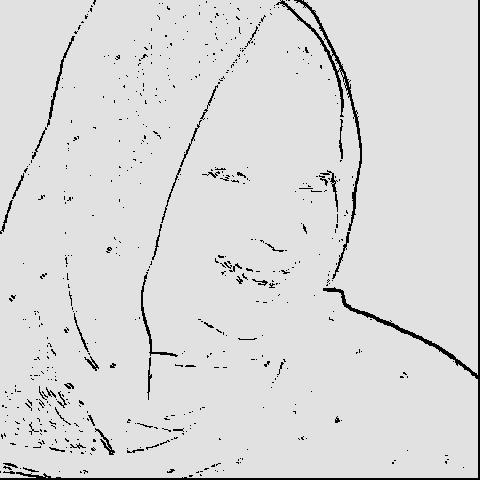}\title{c}&\includegraphics[width=3cm]{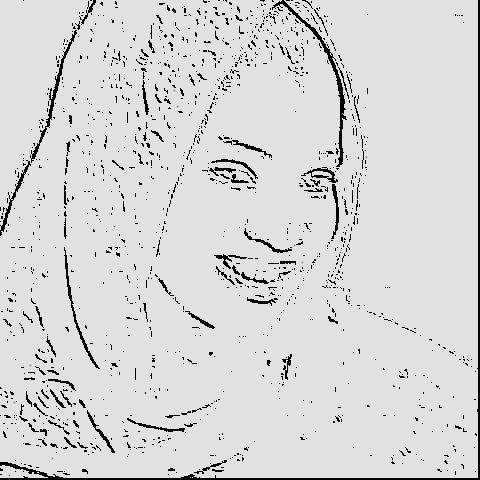}\title{d}&\includegraphics[width=3cm]{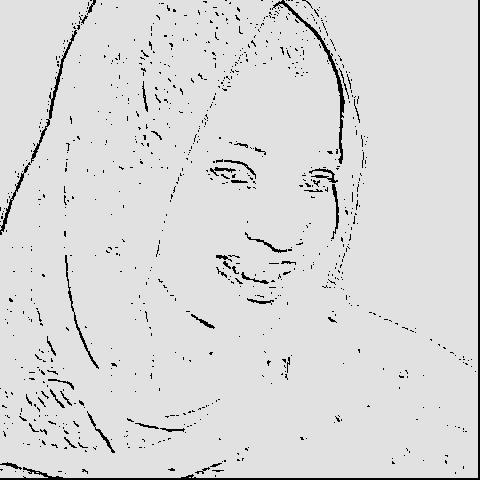}\title{e}\\
\includegraphics[width=3cm]{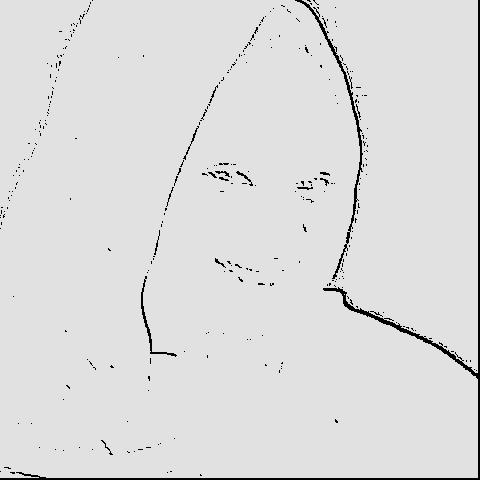}\title{f}&\includegraphics[width=3cm]{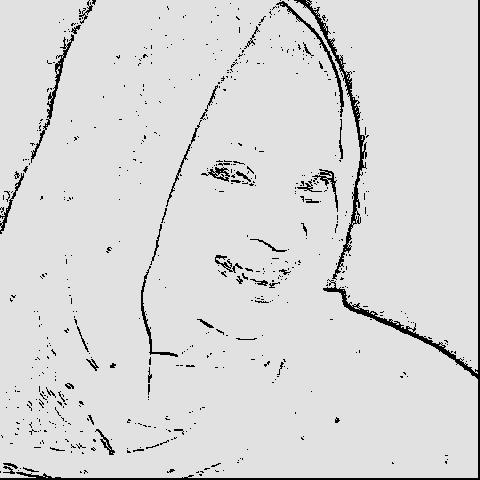}\title{g}&\includegraphics[width=3cm]{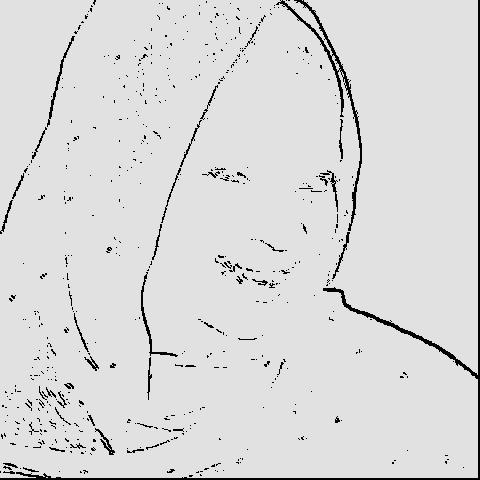}\title{h}\\
\includegraphics[width=3cm]{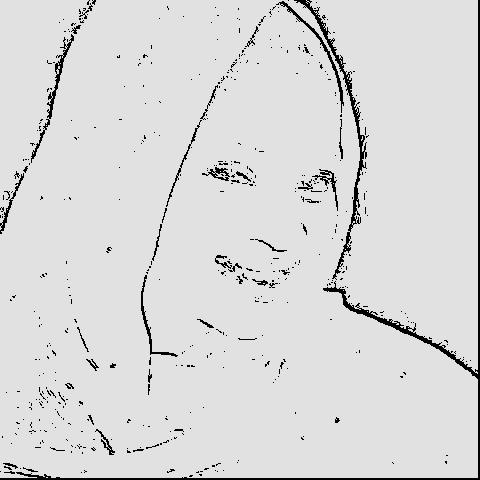}\title{i}&\includegraphics[width=3cm]{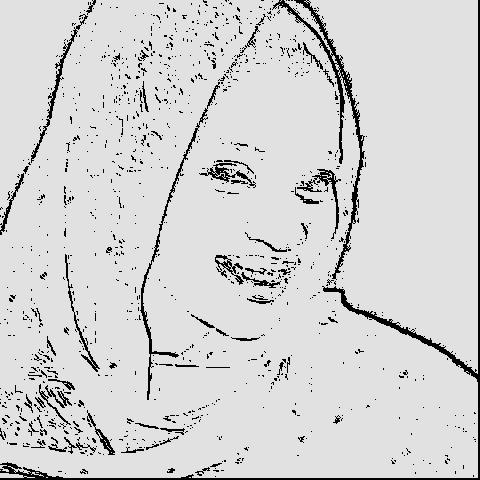}\title{j}&\includegraphics[width=3cm]{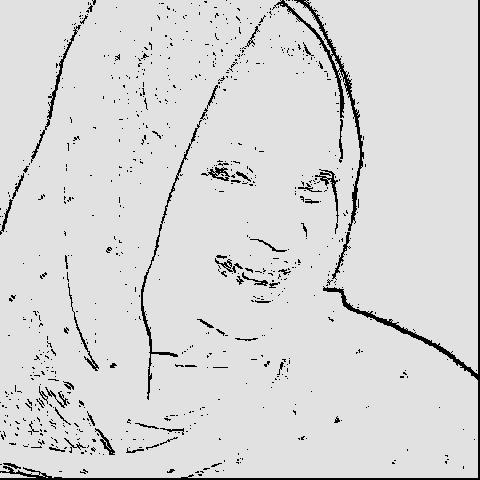}\title{k}\\
\end{tabular}\\
\title{Fig 1: Results facial image segmentation}
\end{center}

In this part of the work we will make a brief presentation of the classification of an image. A presentation of the transformation with the corresponding divergence and the optimal threshold.
The color image segmentation results is that :(a) coresponds to the original image   , (b) image  $D_{ABS}^{(1,0)}$-divergence with k=2.4, (c) image $D_{ABS}^{(1,1)}$-divergence with k=1354, (d) image  $D_{ABS}^{(1/2,-1)}$-divergence and k=2.5, (e) image $D_{ABS}^{(1/2,-1)}$-divergence with k=3.4, (f) image  $D_{ABS}^{(1,0)}$-divergence with k=3.5, (g) image $D_{ABS}^{(1/2,1/2)}$-divergence with k=4.5, (h) image  $D_{ABS}^{(1,1)}$-divergence with k=1350, (c) image $D_{ABS}^{(1,1)}$-divergence with  k=1354, (i) image $D_{ABS}^{(1/2,1/2)}$-divergence with k=5.5, (j) image $D_{ABS}^{(1/2,1)}$-divergence with k=50, (k) image $D_{ABS}^{(1/2,1)}$-divergence with k=80.\\

From these figures (Fig 1), we can observe the experimental  results of our algorithm on a facial image. Using on divergences in the proposed algorithm, we can observe the separation into two classes of our image according to the metric used with an adequate theshold. We can see some differences between the segmentation results images (b-k). For example, the images (b) and (f) given by the $D_{ABS}^{(1,0)}$ divergence showed a bleary delimitation of the original image (a); for the images (c,e,h,i) we used (resp.) $D_{ABS}^{(1,1)}$, $D_{ABS}^{(1/2,-1)}$, $D_{ABS}^{(1,1)}$, and $D_{ABS}^{(1/2,1/2)}$ divergences they correctly delineates the contour of the image (a); for the images (d,g,j) the $D_{ABS}^{(1/2,-1)}$, $D_{ABS}^{(1/2,1/2)}$, and $D_{}^{(1/2,1)}$ divergences results are relatively homogeneous and do a good work.

\subsubsection{Fruit image segmentation}

 \begin{center}
\begin{tabular}{ccc}

\includegraphics[width=3cm]{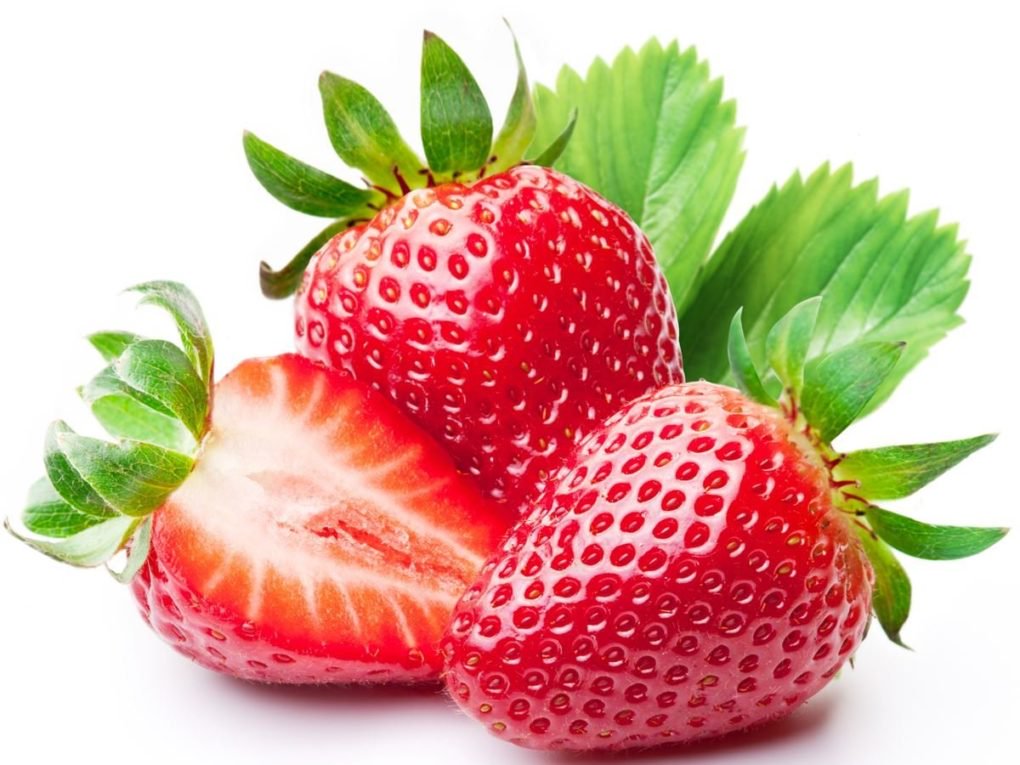}\title{1}&\includegraphics[width=3cm]{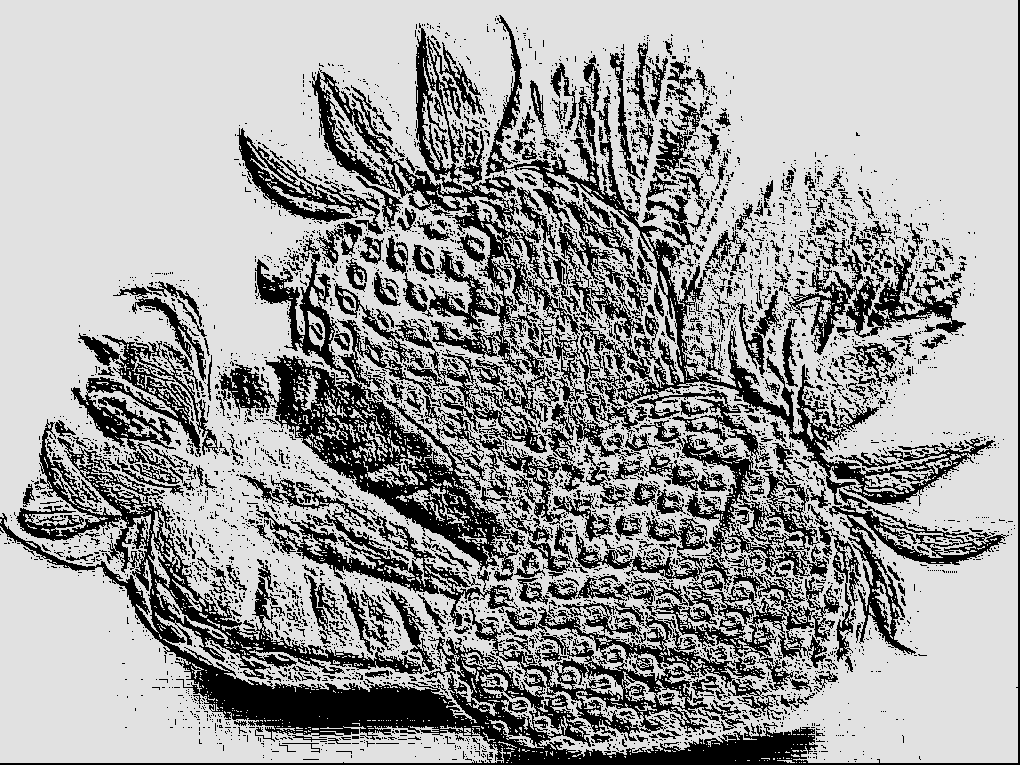}\title{2}\\
\includegraphics[width=3cm]{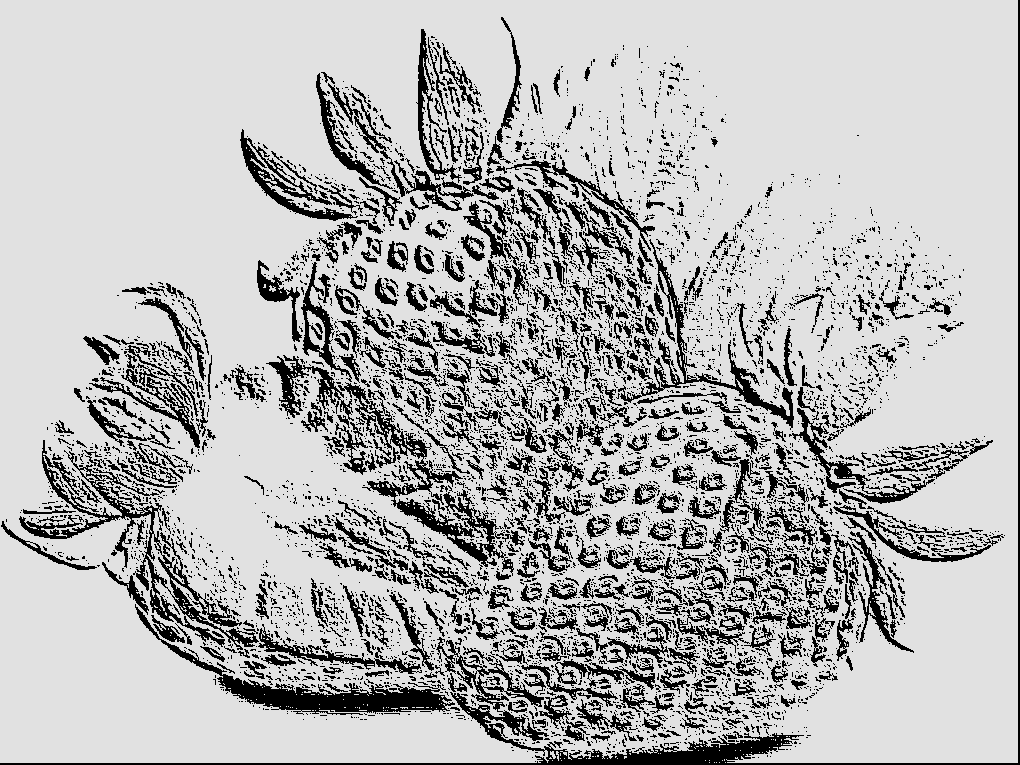}\title{3}&\includegraphics[width=3cm]{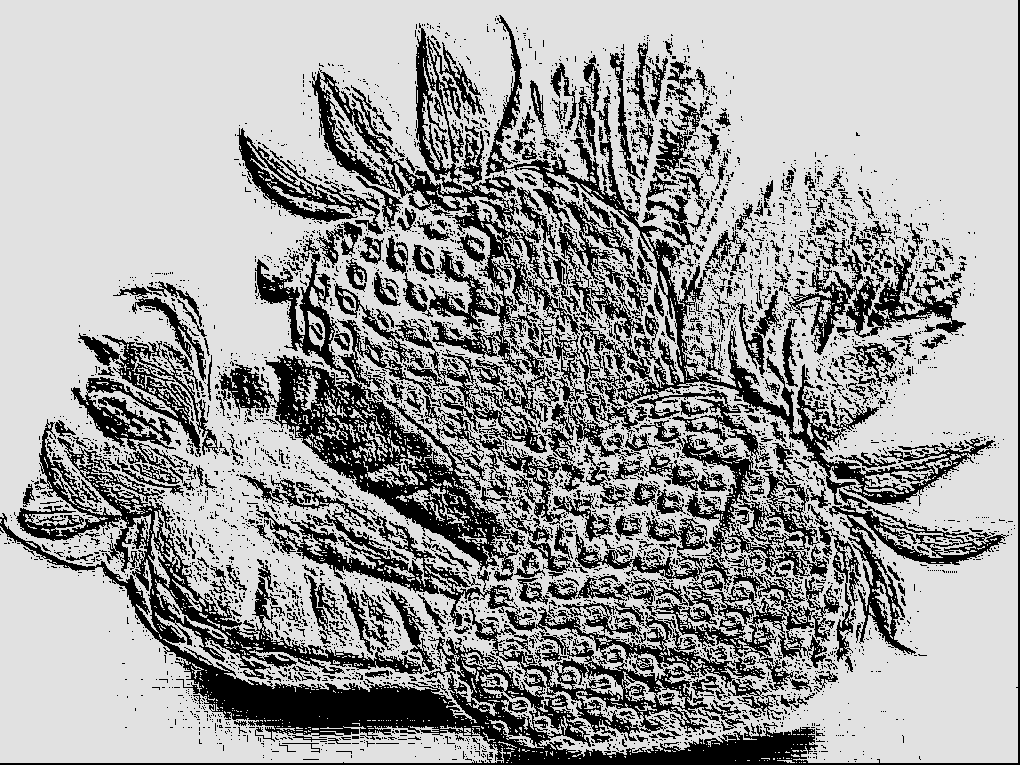}\title{4}&\includegraphics[width=3cm]{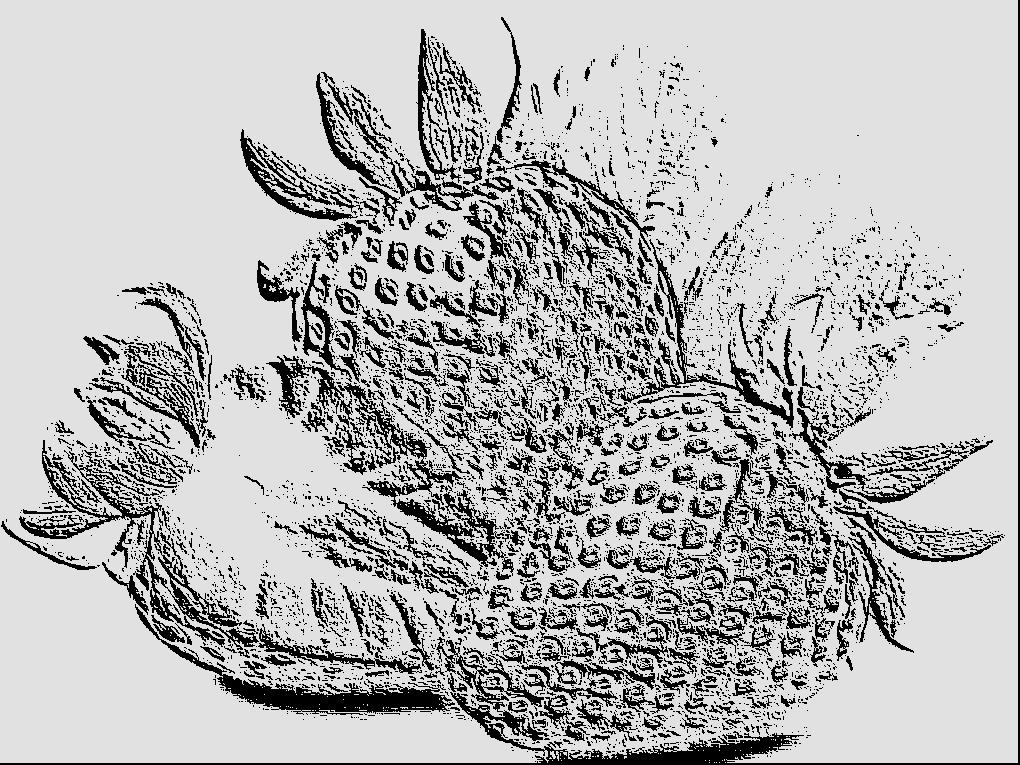}\title{5}\\
\includegraphics[width=3cm]{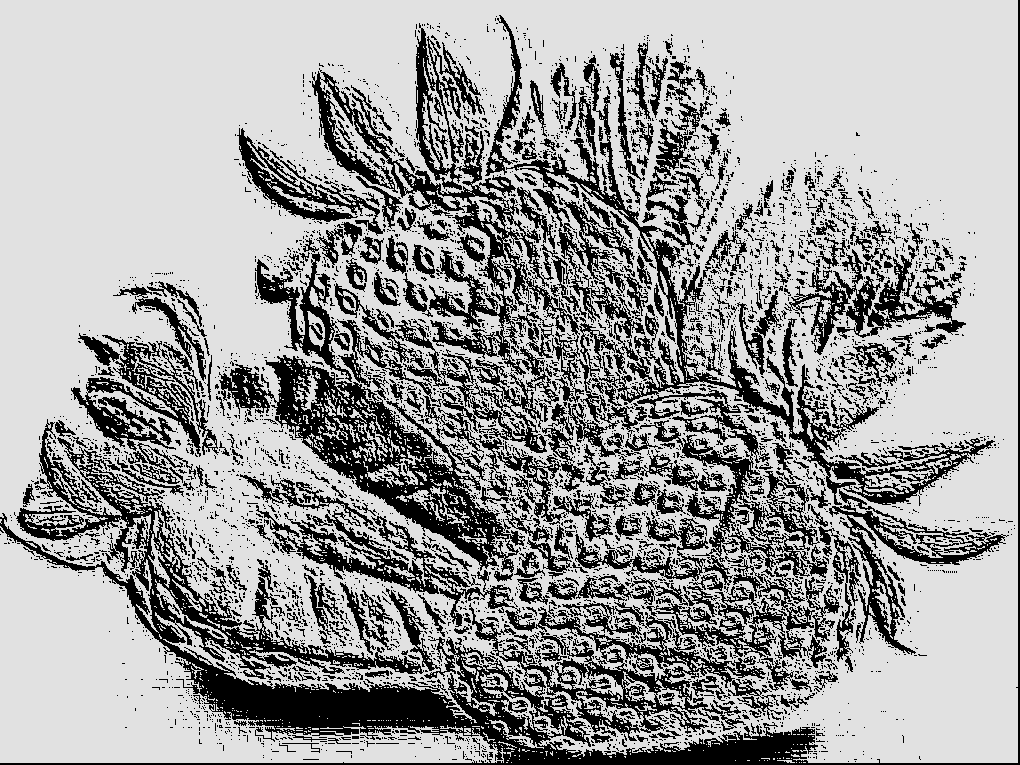}\title{6}&\includegraphics[width=3cm]{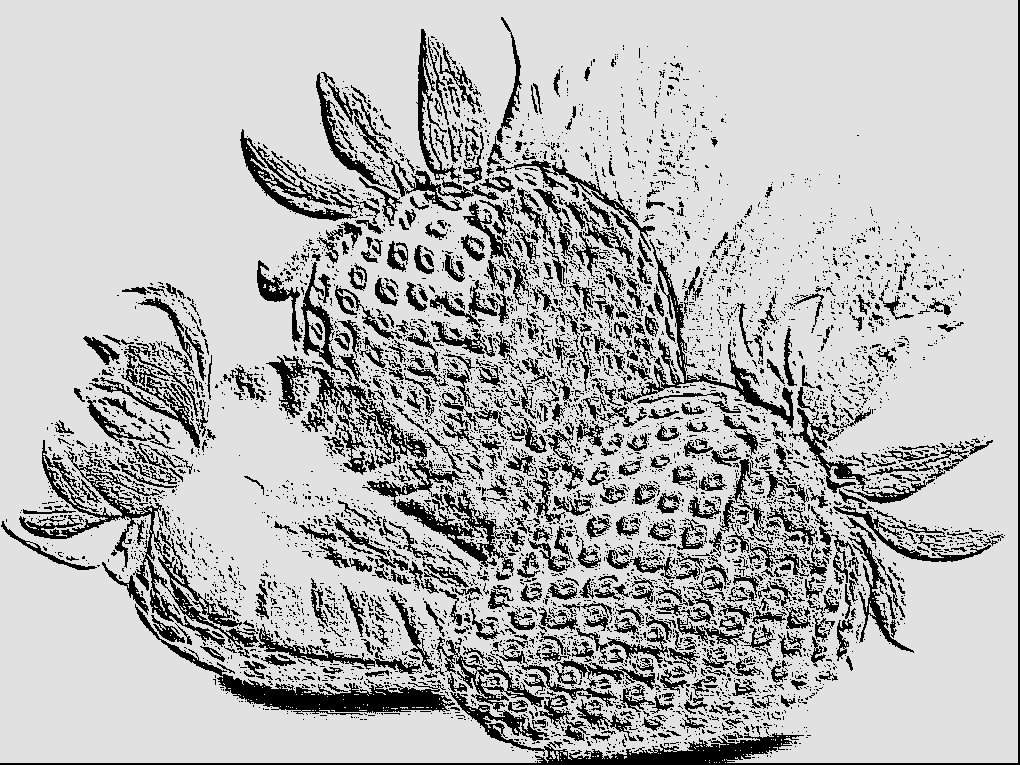}\title{7}&\includegraphics[width=3cm]{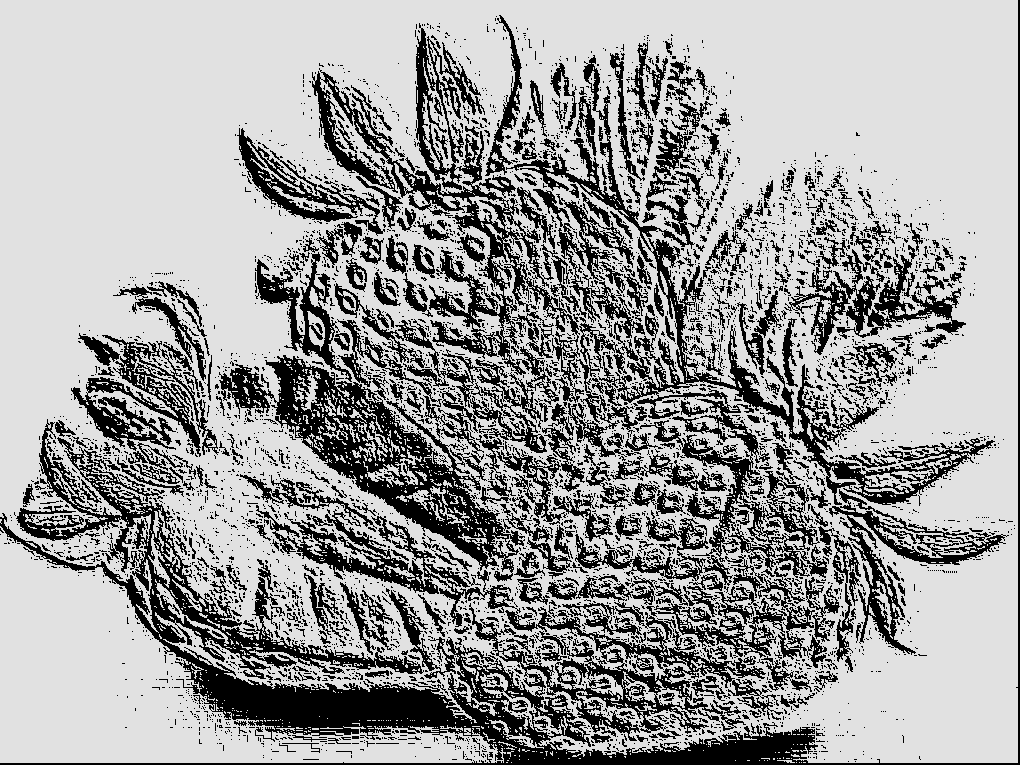}\title{8}\\
\includegraphics[width=3cm]{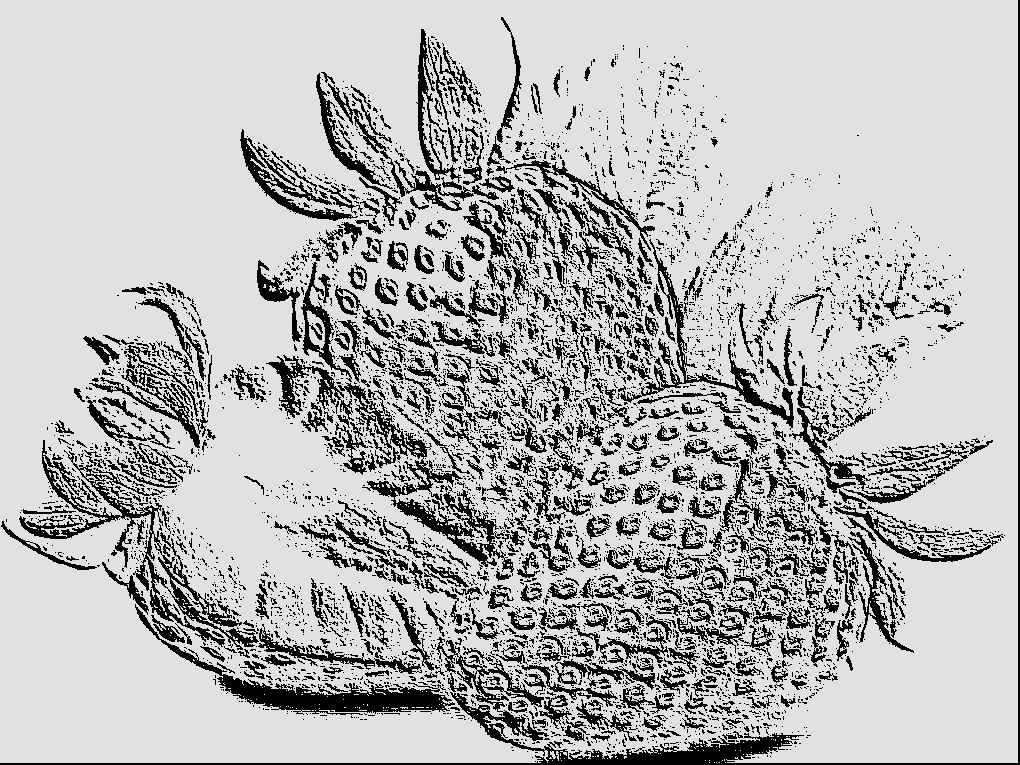}\title{9}&\includegraphics[width=3cm]{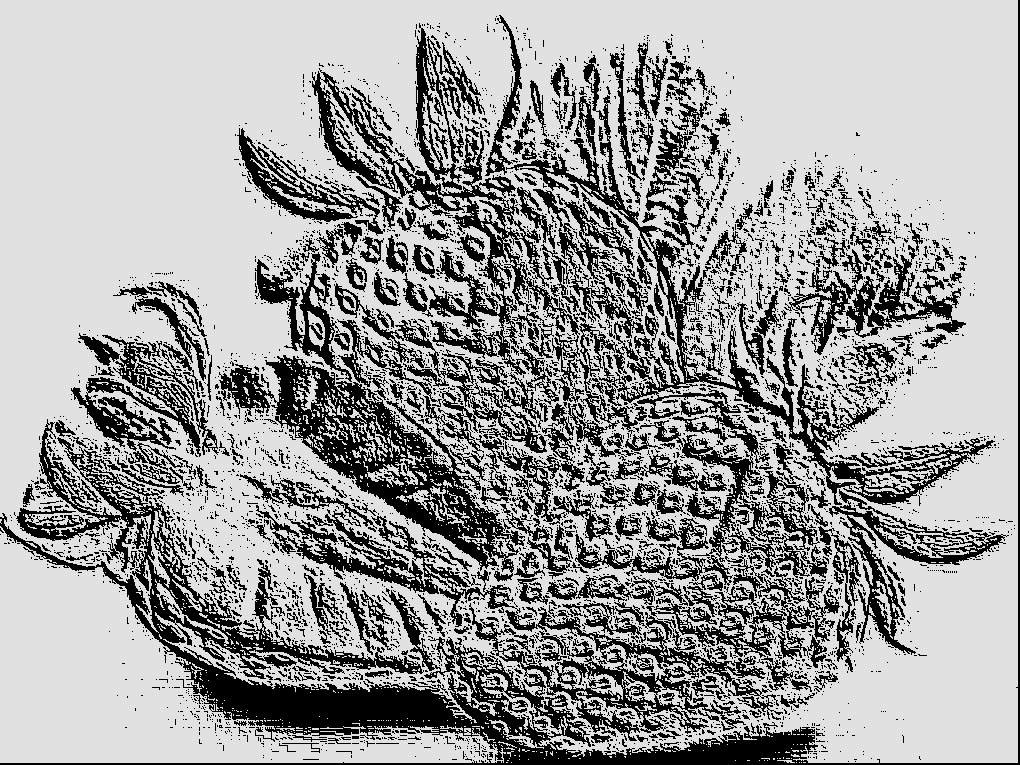}\title{10}&\includegraphics[width=3cm]{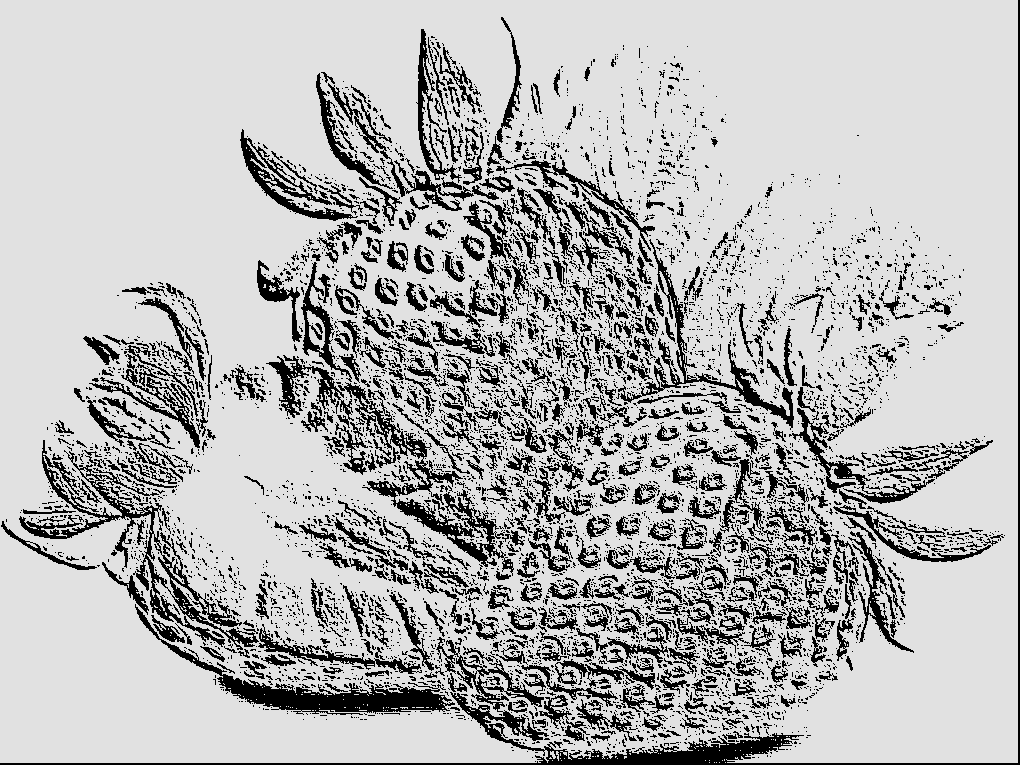}\title{11}\\
\end{tabular}\\
\title{Fig 2: Fruit image segmentation}
\end{center}

The images in Figure 2 correspond to a classification using the proposed algorithm. The classification uses our divergences with a certain decision threshold k. Image 1 corresponds to the original images. The image 2 corresponds to the classification using the divergence $D_{ABS}^{(1,1)}$ with a threshold k = 0.1, the image 3 is that of this divergence with the threshold k = 0.5. The images 4 and 5 correspond to the classification the divergence $D_{ABS}^{(\frac{1}{2},\frac{1}{2})}$ with the thresholds k = 0.1 and k = 0.5 respectively. The classification made with the divergence $D_{ABS}^{(\frac{1}{2},1)}$ with the threshold k = 0.1 and k = 0.5 respectively corresponding to the 6 and 7.The images 8 and 9 are obtained with the divergence $D_{ABS}^{(\frac{1}{2},-1)}$ with the thresholds k = 0.1 and k = 0.5. The images 10 and 11 correspond to the classification with the divergence$ D_{ABS}^{(1,0)}$ with k = 0.1 and k = 0.5 respectively.

From these figure (fig:2), we can observe the experimental results of our algorithm on a fruit image. Using on divergences in the proposed algorithm, we can observe the separation into two classes of our image according to the metrics used with an adequate theshold. For k=0.1 figures (2,4,6,8,10) present the better representation image than k=0.5 images (3,5,7,9,11). The fruits images reprensentations depend two thing k and the diverggences. But in this cas we have a few differences between our divergences.

\section{Conclusion}
A general method to build Hilbertian metrics on probability measures from Hilbertian on $\mathbb{R}_{+}$ was presented. Using results from Cichocki and Amari from the Alpha-Beta-divergence, then we generalized this framework 
by incorporating the symmetry property. We propose a new variant of Alpha-Beta-Symmetric divergence metrics  (ABS-divergence) and kernels associed. Our main contributions consist of, first is  to construct a new family of metrics, ABS-divergence and kernels, and second is to be integrated into SVM and algorithm classification. 
Our results, which are based on a choice of ABS divergence parameters leading to  symmetric kernel $\mathcal{K}_t$, are very efficient compared to classical $\mathcal{K}$-based methods.

\section{Acknowledgments}
This reseach was supported, in part, by grants from NLAGA project, "Non linear Analysis, Ceometry and Applications Projet". ( Supported by the University Cheikh Anta Diop UCAD ).
\newpage




\begin{thebibliography}{99}
\bibitem{23} Andrzej Cichocki , Sergio Cruces , and Shun-ichi Amari, \emph{Log-Determinant Resivited: Alpha-Beta and Gamma log-det divergences}, Entropy, vol. 17, pp. 2988-3034, 2015.
\bibitem{24}{Andrzej Cichocki , Sergio Cruces , and Shun-ichi Amari } Generalized Alpha-Beta Divergences and Their Application to Robust Nonnegative Matrix Factorization.Entropy,vol. 13, pp. 134-170, 2011. 
\bibitem{Cd202} Asa Ben-Hur and Willian Stafford Noble, \emph{Kernel Methods for Predicting protein-protein-interactions}vol. 21, pp. i38-i46, 2005.
\bibitem{Cd98} B. Fr\'enay and M. Verleysen, \emph{Parameter-free kernel extreme  learning for non-linear support vector regression}, Neurocomputing, vol. 74, pp. 2526-2531, 2011.
\bibitem{Cd97} B. Fuglede, \emph{Spirals in Hilbert space}, With an application in information theory, Expositiones Mathematicae, vol. 23, pp. 23-45, 2005.
\bibitem{Cd108} B. Scolkeropf and A. Smola, \emph{Learning with kernels}, MIT Press, Cambridge, MA, 2002.
Bharath K. Sriperumbudur and al, \emph{Hilbert space embeddings and metrics on probability measures}, Journal of Machine Learning Rsearch, vol. 11, pp. 1517-1561, 2010.

\bibitem{Cd105} B. Fuglede, \emph{Spirals in Hilbert space. With an application in information theory }, TO appear in Expositiones Mathematicae (2004).
\bibitem{Cd200} Christopher M.Bishop, \emph{Pattern Recognition and Machine Learning}, Springer 2006.
\bibitem{Cd202} D. B. Thiyam and al, \emph{Optimization of alpha-beta log-det divergences and their application in the spatial filtering of two class motor imagery movements}, Entropy, vol. 19, pp. 89, 2017.
\bibitem{Cd201} D. Olszewski and B. Ster, \emph{Asymmetric clustering using the alpha-beta divergence}, Pattern recognition, vol. 47, pp. 2014-2041, 2014.
\bibitem{Cd95} F. Tops$\phi$e, \emph{Jenson-shannon divergence and norm-based measures of discrimination and variation},Preprint (2003).
\bibitem{Cd456} Ho Chung Leon Law, Dougal J. Sutherland, Dino Sejdinovic and Seth Flanman, \emph{Bayesian Approches to distribution regression}, Proceedings of the $21^{st}$ International Conference on Artificial Intelligence and Statistics (AISTATS), Lamzorote, Spain. PMLR, vol. 84, 2018. 
\bibitem{Cd96} I.J. Schoenberg, \emph{Metric space and positive definite function}, Trans. Amer. Math.Soc,vol.44,pp. 522-536, 1938.
\bibitem{Cd656} F. Amara, M. Fezari and H. Bouboura, \emph{An improved GMM-SVM system based on distance metric for voice pathology detection}, Applied Mathematics and Information Sciences An International Journal, vol. 10, N. 3, pp. 1061-1070, 2016.
\bibitem{Cd43} I.W. Sumarjaya, \emph{A servey of kernel-type estimators for copula and their applications}, Journal of physics, conf.series 893, 012027, 2017.
\bibitem{Cd99} J. Lafferty and G. Lebanon, \emph{Diffusion kernels on statistical manifolds}, Journal of machine learning Research, vol. 3175, pp. 129-163, 2005. 

\bibitem{Cd104} J. P. R. Christensen C. Berg and P. Ressel, \emph{Harmonic Analysis on Semigroups}, Springer, New York , vol. 29, pp. 438, 1984.
\bibitem{Cd109} John Shawe-Taylor and Mello Cristianini, \emph{Kernel Methods for Pattern Analysis} 2004.
Krikamol Muandet and al, \emph{Kernel mean embedding of distribution}, A review and Beyond, Foundations and trends in Mchine Learning, vol. 10, N. 1-2, pp. 1-141, 2017.
\bibitem{Cd102} M. Hein, T. N. Lal, and O. Bousquet, \emph{Hilbertian metrics on probability measures and thier application in SMVs}, Acceptide at DAGM Springer , vol. 3175 , pp. 270-277, 2004.

\bibitem{Cd106} M. Hien, O. Bousquet, and B. Scholkeropf, \emph{Maximal margin classification for metric spaces}, Journal of Computer and System Sciences, vol. 71, pp. 333-359, 2005.

\bibitem{Cd94} O. Chappelle, P. Haffiner, and V. Vapnik, \emph{SVMs for histogram-based image classification}, IEEE Transaction on Neural Networks,  vol.   10, pp. 1055-1064, 1999.
\bibitem{Cd101} P. J. Moreno,P. P. Hu, and Vasconcelos, \emph{A Kullback-Leibler divergence based kernel for SVM classification in multimedia application}, NIPS, vol. 16 ,2003.
\bibitem{Cd100} T. Jebara and R. Kondor, \emph{Bhattacharyya and exppected likelihood kernels}, In 16th Annual Conference on Learning Theory (COLT), pp.  57-71, 2003.
\bibitem{Cd101} T. J. Abrahamsen , L. K. Hansen and O. Winther, \emph{Kernel Methodes for Machine Learning}, Technical University of Denmark (DTU), PHD, N. 299, 2013.
\bibitem{Cd107} V. Vapnik , \emph{ Statistical Learning Theory}, Wiley, New York 1998.


\end{thebibliography}
\end{document}